# Antenna-enhanced Optoelectronic Probing of Carbon Nanotubes

*Nina Mauser (1), Nicolai Hartmann (1), Matthias S. Hofmann (2), Julia Janik (1), Alexander Högele (2), Achim Hartschuh (1)\**

(1) Department Chemie und Center for NanoScience (CeNS), Ludwig-Maximilians-Universität, 81377 München, Germany

(2) Fakultät für Physik und Center for NanoScience (CeNS), Ludwig-Maximilians-Universität, 80539 München, Germany

*E-mail: achim.hartschuh@cup.uni-muenchen.de

**We report on the first antenna-enhanced optoelectronic microscopy studies on nanoscale devices. By coupling the emission and excitation to a scanning optical antenna, we are able to locally enhance the electroluminescence and photocurrent along a carbon nanotube device. We show that the emission source of the electroluminescence can be point-like with a spatial extension below 20 nm. Topographic and antenna-enhanced photocurrent measurements reveal that the emission takes place at the location of highest local electric field indicating that the mechanism behind the emission is the radiative decay of excitons created via impact excitation.**

During the last two decades rapid progress in the research field of new nanomaterials and technologies to complement conventional silicon systems has been made. A variety of electronic and optoelectronic devices that involve different nanostructures have been realized, such as field-effect transistors, photovoltaic and light-emitting devices often based on carbon as active component[1,2]. Due to the size-mismatch between the wavelength of visible light and the dimensions of these devices the full potential of the nanomaterial could not be exploited in optoelectronic applications so far. For the same reason the details of the electrical-to-optical and optical-to-electrical transduction mechanisms remain hidden in conventional optical microscopy.

In the radio frequency range this size-mismatch is compensated by macroscopic antenna structures since more than a century. Optical antennas are metallic nanostructures capable of converting free-propagating visible radiation into localized energy and vice versa[3]. Using these phenomena the performance of nanoscale optoelectronic devices could thus be improved and, if the antenna is scanned across the sample in close proximity, high-resolution images could be obtained as suggested in ref. 3. Static antenna configurations have been realized already for a variety of sample materials including photovoltaic devices, OLEDs and pn-junctions[4-9]. Scanning optical antennas can enhance different photophysical processes at a chosen position, outlined in Figure 1 but have almost exclusively been used for Raman and photoluminescence spectroscopy (Figure 1a) [10-12]. In this application the antenna enhances both the excitation and the emission rate of the investigated material leading to sub 20 nm spatial resolution and enhanced detection sensitivity.

Photocurrent and electroluminescence microscopy (Figure 1b and c) are complementary techniques probing the absorption and emission of light, respectively, not necessarily involving the same sample states. The information that can be obtained is thus inherently different, comparable to the differences between absorption and fluorescence spectroscopy. Moreover both photocurrent and electroluminescence signals can each have diverse physical origins as discussed below. In photovoltaic devices only the excitation rate is enhanced by the optical antenna following Figure 1b. The increased absorption cross section leads to larger photocurrents, locally increasing the conversion efficiency of photovoltaic devices[13]. Recently, we applied antenna-enhanced photocurrent microscopy to CNTs and resolved their photovoltaic response on a length scale of 30 nm. Following the reciprocity theorem optical antennas may not only increase the efficiency of light absorption but also of light emission (Figure 1c). This can be exploited to enhance light emission in LEDs, for example[3,4]. Antenna-enhanced microscopy could thus have the potential to visualize electroluminescence (EL) intensities and energies on a sub-diffraction length scale for the first time.

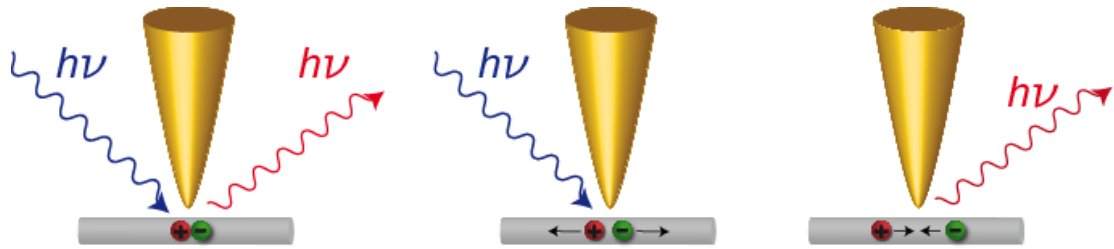

**Figure 1.** Applications of antenna enhancement: a) Optical spectroscopy, b) Photovoltaics and c) Electroluminescence (adapted from ref. 3).

In this article we present the realization of all three scanning antenna schemes outlined in Figure 1 using carbon nanotubes (CNTs) as sample material. This allows us to obtain a comprehensive description of the observed optoelectronic nanoscale phenomena and address remaining open questions regarding the origin and the spatial extension of the source of electroluminescence.

To demonstrate all three antenna schemes and to illustrate their capabilities for the non-destructive functional characterization of materials we fabricated a carbon nanotube based device consisting of a heterogeneous network structure from which the different types of signals can be obtained (see Supporting Information Figure S1 for an overview of the network structure). We first studied the device by confocal laser microscopy, a platform used intensively in the past to gain valuable information on the optoelectronic properties of carbon based devices. Different mechanisms behind the electroluminescence of single CNTs including radiative charge carrier recombination due to ambipolar transport[14-16], impact excitation[17,18], Joule heating[19,20] and phonon-assisted decay[21] have been discussed. In principle all of these mechanisms could contribute to the EL in a given device to a varying degree. Photocurrent microscopy has been used to answer questions concerning the Schottky barrier formation at the CNT-metal contacts, potential modulations and pn-junctions along CNTs[22-25].

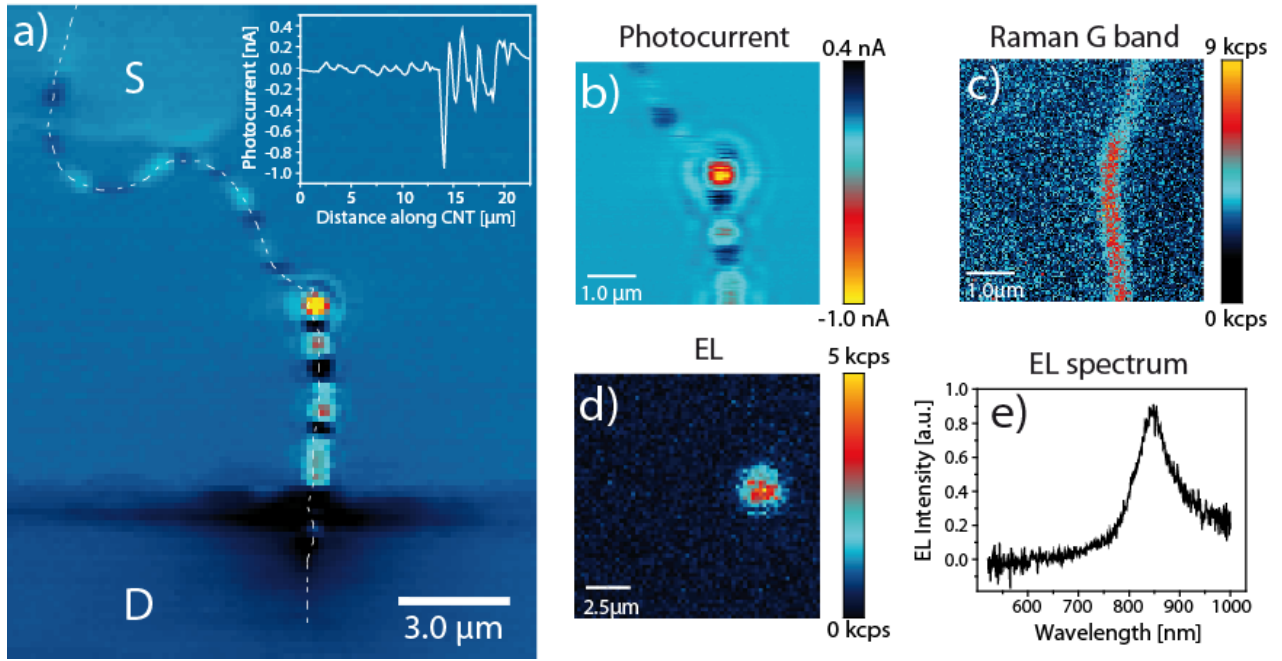

**Figure 2.** Confocal study. a) Confocal zero bias photocurrent image of a CNT network on glass. S and D denote source and drain electrode. The inset shows the zero bias photocurrent signal detected along the white dashed line. b) Confocal zero bias photocurrent image (zoom). c) Simultaneously taken confocal Raman G band image of the same region as in b). d) Confocal electroluminescence image. **e)** Electroluminescence spectrum recorded at the bright EL spot in d).

Figure 2 presents the characterization of the investigated device by means of conventional confocal photocurrent, Raman and electroluminescence microscopy. The device consists of a source (S) and drain (D) electrode made of gold fabricated by optical lithography with a separation of 12 μm that is bridged by CNTs. The CNTs were grown on a quartz substrate by chemical vapor deposition using an iron-ruthenium catalyst with a hydrogen/methane gas mixture at a temperature of 850°C [26]. They were contacted by two electrodes consisting of 0.5 nm titanium / 30 nm gold fabricated via optical lithography and a standard lift-off process. The optical setup is an inverted oil-immersion microscope with high numerical aperture (1.49) and a He-Ne laser (632.8 nm) as excitation source that can be used either for Raman, photocurrent or electroluminescence measurements. For antenna-enhanced measurements it is combined with a shear-force tuning fork AFM using a solid gold tip as a probe. More experimental details can be found elsewhere[13].

A scanning confocal zero bias photocurrent image of the device is presented in Figure 2a and a zoom in Figure 2b showing typical features such as photocurrent fluctuations along the CNT caused by local built-in electric fields[13,24]. The inset illustrates the photocurrent signal along the channel following the white dashed line. The sudden increase in photocurrent by a factor of about 10 in the middle of the channel can be explained with the help of the Raman measurement displayed in Figure 2c that was simultaneously taken with the photocurrent measurement shown in Figure 2b. In contrast to the region of strong photocurrent fluctuations there is no detectable Raman G band signal in the region of weak photocurrent signals indicating an off resonance condition of the excitation energy for this specific CNT. We therefore conclude that more than one CNT of different chiralities is responsible for the current transport.

The electroluminescence image presented in Figure 2d is obtained by switching off the excitation light and applying a source-drain bias of 9 V. Bright EL appears approximately at the location of the sudden change in the photocurrent and Raman signal. The emission has a spectral maximum in the near-infrared at about 850 nm (Figure 2e). The spatial resolution of the photocurrent and the Raman image is diffraction-limited and given by approximately half of the laser wavelength (~300 nm). Note the reduced spatial resolution of ~1.6 μm in the electroluminescence image that reflects the effective size of the photodetector since no additional pinhole was used in the detection path. The optimum spatial resolution of few hundred nanometers achievable using focused light is not sufficient to resolve the size of the source of electroluminescence and to correlate the observed signals with the nanotube network structure.

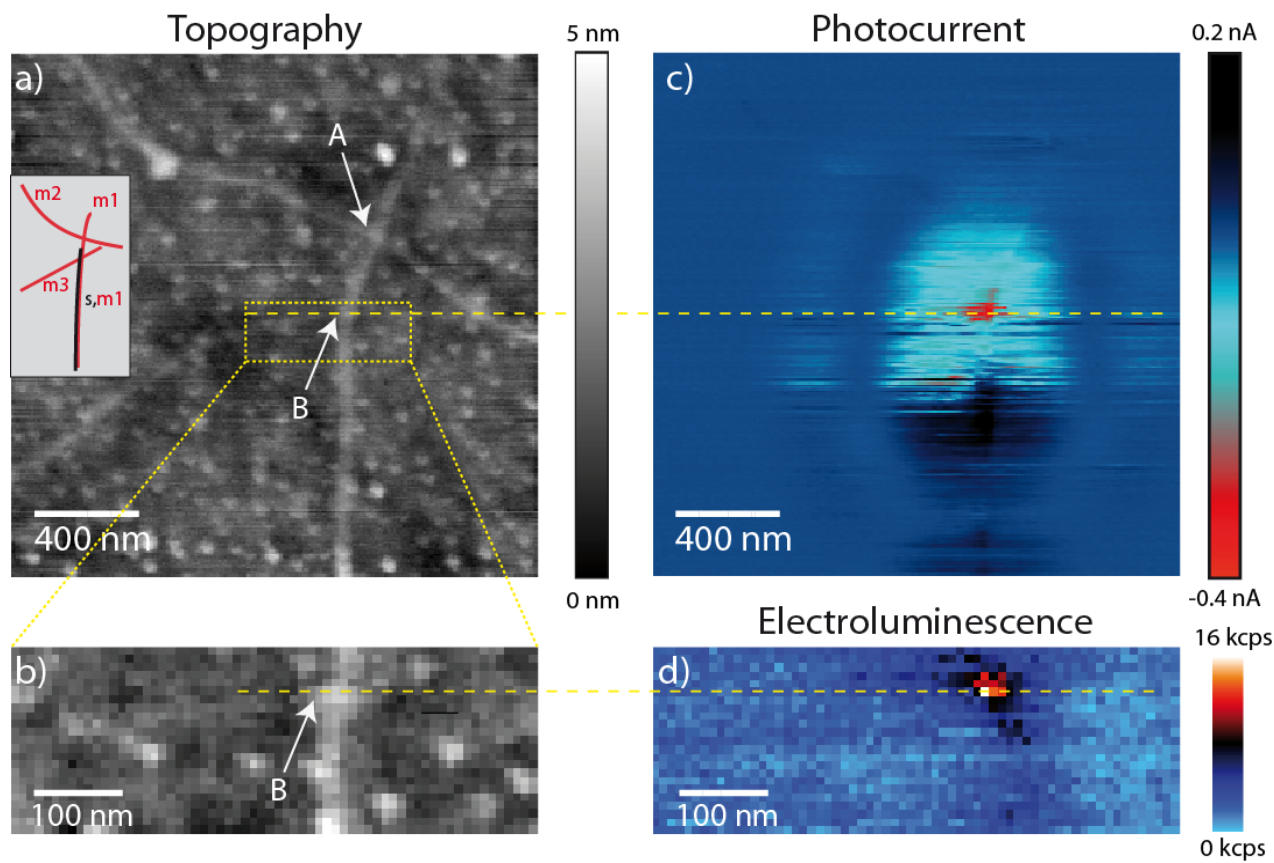

**Figure 3.** Antenna-enhanced optoelectronic study. a) Topography image revealing two CNT-CNT junctions denoted by A and B. Inset: sketch of the region containing different metallic (m) and semiconducting (s) CNTs. The corresponding Raman maps are shown in the Supporting Information Figure S1a and b. b) Magnified view of the topography image in the area of the lower junction B. c) Antenna-enhanced zero bias photocurrent image. d) Antenna-enhanced electroluminescence image. Sub-diffraction resolution is clearly achieved for both signals. The location of strongest absolute photocurrent and electroluminescence coincide exactly with the position of junction B as indicated by the horizontal dashed lines.

In the following we show that antenna-enhanced optoelectronic microscopy can reach a spatial resolution as high as 40 nm. Using this technique we are able to record sub-diffraction Raman, photocurrent and electroluminescence images of our device.

The topography images (see Figure 3a and 3b) of the region where EL was observed reveal the appearance of two CNT-CNT junctions denoted by A and B. The simultaneously taken antenna-enhanced photocurrent image is shown in Figure 3c. By increasing the light absorption cross section in a nanoscale volume given by the tip apex, an enhanced narrow signal in addition to the broad confocal background observed by focused light in Figure 2a and 2b is obtained, that shows a strong maximum followed by a rapid change of signal sign. The high spatial resolution clearly reveals that junction B is responsible for the large photocurrent signal. Since the photocurrent originates from the dissociation of electron-hole pairs a strong local electric field exists at this position and a direct proportionality between the electric field and the photocurrent is assumed[22-25]. We expect that the electric field is caused by the crossing of two or three nanotubes forming a Schottky contact consisting of at least one metallic and one semiconducting CNT as indicated by Raman measurements presented in the Supporting Information Figure S1 and S2 and Note 1. The formation of a local electric field at the junction of a metallic and a semiconducting CNT has been demonstrated before by transport and photocurrent measurements[27,28].

The electroluminescence measurement of the device reveals a single bright emission site (Figure 3d). The observed enhanced emission appears point-like in good contrast to the confocal background and is localized at the position of the strongest electric field probed by the photocurrent configuration at junction B discussed above. Note that we do not observe any emission at junction A. We attribute this to a weaker electric field than at junction B due to the metallic character of both crossing CNTs (see Supporting Information Note 1 and Figure S1 and S2). The antenna-enhanced EL measurement makes it for the first time possible to determine the spatial extension of the emitting nanotube segment. In the past, the extension of the emission site was hidden behind the broad diffraction-limited signal. In the case of ambipolar transport this figure gives information about the spatial extent of the recombination length and has been determined in ref. 15 to be equal to or less than 2 µm. In the case of impact excitation the size of the light source is a figure of merit for the screening length[17]. By taking cross sections in the directions perpendicular to and along the CNT we determine the spatial extension of the emission site.

Figure 4a shows the cross section perpendicular to the nanotube. The narrow signal (highlighted by the red line) that is due to the near-field interaction between the emission and the antenna has a spatial width of about 45 nm ± 10 nm (full width at half maximum). Since the diameter of the CNT of about 1 nm is substantially smaller, the width of the recorded signal can be taken as the spatial resolution of the experiment The signal background consists of two further contributions that can be seen in the insets in Figure 4a and b: The nearly uniform confocal background resulting from diffraction-limited collection that appears in blue; and an additional contribution seen as black area with a spatial width of around 100 nm that could be attributed to a long-range electrostatic interaction between tip and nanotube. During the scanning process the gold tip might get slightly electrostatically charged and act as a weak local gate influencing the band energy of the nanotube. Such a Coulomb interaction will scale with the inverse of the tip-sample distance and is thus expected to lead to a longer ranged interaction and hence broader signal contrast compared to the near-field optical interaction that is based on a dipolar interaction[3].

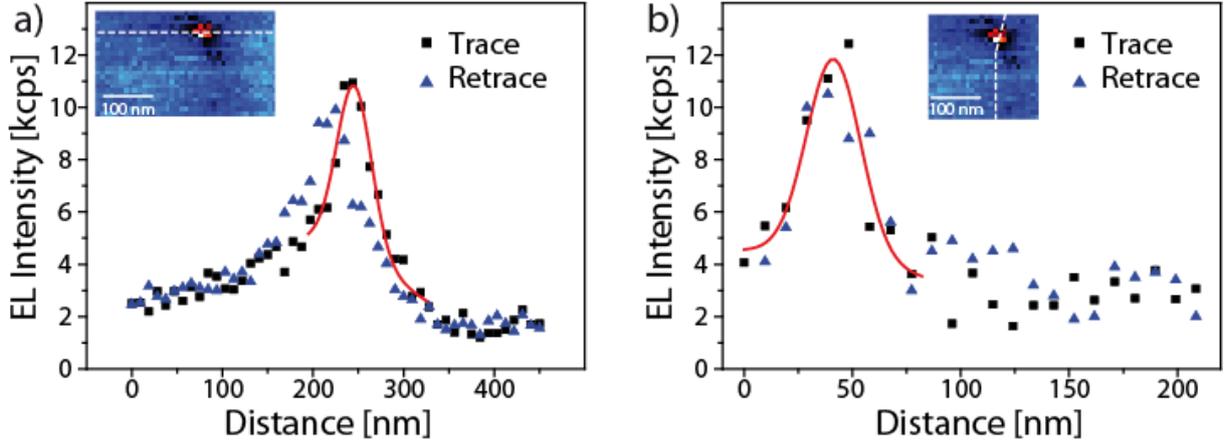

**Figure 4.** Size of the source of electroluminescence. a) Antenna-enhanced EL cross sections perpendicular and b) along the CNT. The red lines are a guide to the eye to highlight the near-field contribution. Since the cross section taken along and perpendicular to the CNT have about the same width of about 40 nm the extension of the EL source along the CNT needs to be substantially smaller, probably below 20 nm.

The cross section taken along the CNT (Figure 4b) shows a spatial width of about 30 ± 10 nm. The values for both directions lie in the expected range of the spatial resolution of our system that is determined by the tip size. Since the signal width recorded along the nanotube is similar to the spatial resolution determined from the perpendicular cross section as discussed above, the actual width of the EL source must be substantially smaller[29]. We therefore conclude that the emission is point-like and occurs on a length scale smaller than 20 nm.

From Figure 4a the enhancement factor f of the radiative rate $f_{rad} = k_{rad}^{antenna} / k_{rad}$ induced by the antenna can be estimated. Neglecting changes in the non-radiative relaxation rate $f_{rad}$ is equal to the intensity enhancement $f_{rad} = I_{NF} / I_{FF} \sim 9$ kcps / 2 kcps, where $I_{NF}$ is the intensity of the enhanced near-field signal, and $I_{FF}$ the intensity of the confocal far-field signal. The present value of $f_{rad} \sim 4.5$ is comparable to previous results obtained for the photoluminescence from carbon nanotubes with etched gold tips[30].

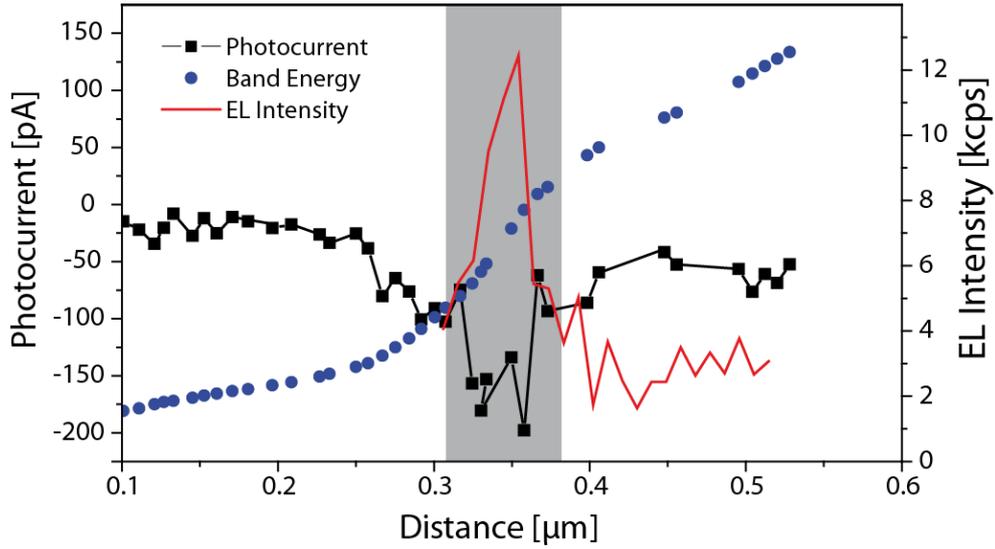

**Figure 5.** Spatial correlation between EL and photocurrent. Antenna-enhanced zero bias photocurrent measured along the device (black square symbols), band energy (blue round symbols) and measured EL intensity (red line). The EL peaks at the position of highest photocurrent and highest band energy gradient.

From Figure 3 it is clear that a spatial correlation between the photocurrent and the electroluminescence exists. Both signals peak at the position at which the vertically oriented CNT carrying the photocurrent crosses a second, horizontally oriented CNT slightly visible in the topography image. Since the photocurrent signal is proportional to the dissociation rate of the optically created excitons, it scales with the local electric field that is present in the excitation region. Spatial integration of the photocurrent along the CNT gives the electron band energy profile $E=-\int I_{PC}dx$ [13] as illustrated in Figure 5. The black squares represent the antenna-enhanced photocurrent obtained by fitting the near-field contribution of the signal cross sections perpendicular to the nanotube by a Gaussian function for all positions along the nanotube. The blue circles represent the electron band energy profile E. The negative photocurrent peak results in a steep increase of the band energy.

By overlaying the photocurrent and electroluminescence signal along the CNT as shown in Figure 5 (grey shaded region), we see that the electroluminescence spot appears at the position of highest (absolute) photocurrent, at the steepest region of the energy band, respectively. This indicates that the electroluminescence in the present device is caused by impact excitation[17,18]. This mechanism requires a strong electric field in which charge carriers are accelerated locally gaining the kinetic energy sufficient for creating excitons via collisions with other carriers. These excitons can decay radiatively generating an electro-luminescence signal with a photon energy that corresponds to the involved exciton energy.

In principle, other mechanisms besides impact excitation could also contribute to the observed emission as noted in the introduction. However, substantial contributions of these mechanisms are unlikely based on our device configuration and our experimental results. More specifically, we can rule out charge carrier recombination due to ambipolar transport for our non-gated devices. CNT devices without gate voltage are known to act as p-doped unipolar devices under ambient conditions[31]. The ambipolar regime could be reached by fabricating a pn-junction[32], asymmetric contacts[33] or suitable bias-gate-voltage combinations[15,16]. We can also exclude phonon-assisted radiative decay from the M-point discussed in ref. 21, since we do not observe the associated emission peaks at 1.4 and 1.8 eV (Figure 2e). Finally, electroluminescence can also be due to radiative recombination from thermally populated higher energy bands due to Joule heating. Joule heating is expected to be strongest at the position of the largest voltage drop along the channel, which could indeed occur at the junction where the electrical resistance will be high. However strong thermal emission was only observed for suspended metallic CNTs[19,20], while for non-suspended CNTs, as in our case, only very weak emission occurred due to highly efficient substrate cooling[21]. Based on this discussion and the occurrence of the EL at the position of the strongest local field as determined from the photocurrent data as well as its spectral characteristics we conclude that impact excitation is the main cause for the observed emission.

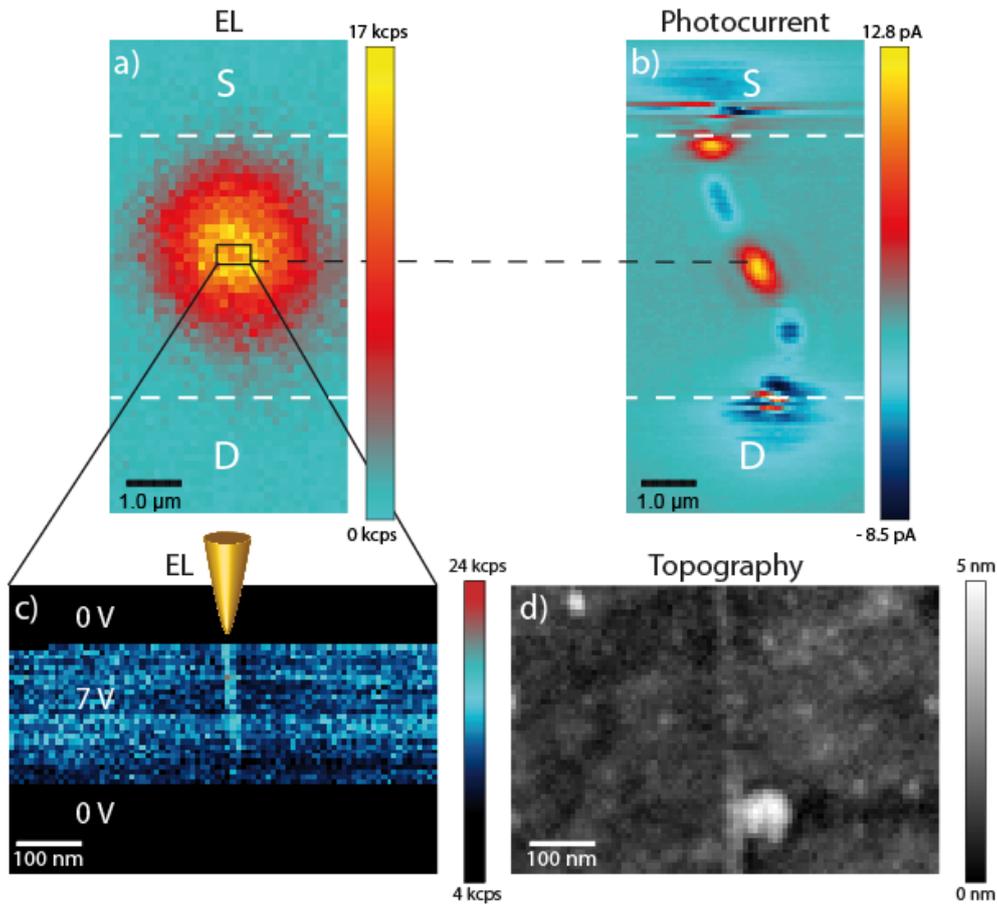

**Figure 6.** Antenna-enhanced EL of a second device. a) Confocal EL image and b) Confocal photocurrent image. The EL occurs at the position of strongest photocurrent signal. c) Antenna-enhanced EL image. The confocal and antenna-enhanced EL appear upon biasing (7 V). d) Simultaneously taken topography image. The high spatial resolution reveals an emitting nanotube segment that is longer than 100 nm.

While we observed pinning of the EL to a point-like region of about 20 nm for the device discussed above, another device showed EL emission from an extended region of more than 100 nm length with weaker contrast (see Fig. 6c). As can be seen from the confocal EL and photocurrent images (Fig. 6a and b), the EL occurs at about half of the channel length at the position of strongest photocurrent. As for the previous device this indicates, that the EL is caused by impact excitation. However, in contrast to the first device, the EL does not appear at a cross-junction but along a single CNT with an extended electric field in this region. Local built-in electric fields along CNT devices have been observed before and can be attributed to structural defects or environmental effects such as charge traps in the substrate or doping by local adsorbants[24].

In summary, we present a comprehensive analysis of the optoelectronic processes in CNT devices using antenna-enhanced Raman, photocurrent and electroluminescence probing. This technique makes it for the first time possible to determine a value for the spatial extension of the EL emission with a spatial resolution of ~ 40 nm. We show that the EL emission can be pinned to a point-like region occurring at a nanotube-nanotube crossing, but can also extend over more than 100 nm. By correlating the photocurrent and the electroluminescence image, we conclude that the emission is due to impact excitation that happens at the position of strongest electric field. We demonstrate antenna-enhanced optoelectronic probing using carbon nanotubes as sample material, but we expect it to be applicable to other materials such as inorganic nanowires as well.

**Acknowledgements**:

We thank Philipp Altpeter for his help with the optical lithography. Financial support by the ERC through the starting grants NEWNANOSPEC and QuantumCANDI, and the Deutsche Forschungsgemeinschaft (DFG) through the Nanosystems Initiative Munich (NIM) are gratefully acknowledged.

**Supporting Information:**

Additional Information on the CNT network and its characterization of the network via Raman spectroscopy. This material is available free of charge via the Internet at http://pubs.acs.org.

1. Avouris, Ph.; Chen, Z.; Perebeinos, V. *Nature Nanotech.* **2007**, *2*, 605-615.

2. Nanot, S.; Hároz, E. H.; Kim, J.-H.; Hauge, R. H.; Kono, J. *Adv. Mater.* **2012**, *24*, 4977-4994.

3. Bharadwaj, P.; Deutsch, B.; Novotny, L. *Adv. Opt. Photon.* **2009**, *1,* 438-483.

4. Atwater, H. A.; Polman, A. *Nature Mater.* **2010**, *9,* 205-213.

5. Catchpole, K. R.; Polman, A. *Opt. Express* **2008**, *16,* 21793–21800.

6. Nakayama, K.; Tanabe, K.; Atwater, H. A. *Appl. Phys. Lett.* **2008**, *93,* 121904.

7. Liu, C.; Kamaev, V.; Vardenya, Z. V. *Appl. Phys. Lett.* **2005**, *86,* 143501.

8. Pillai, S. et al. *Appl. Phys. Lett.* **2006**, *88,* 161102.

9. Sun, J.; Khurgin, J.; Soref, R. *J. Opt. Soc. Am. B* **2008**, *25,* 1748–1755.

10. Mauser, N.; Hartschuh, A. *Chem. Soc. Rev.* **2014**, *43,* 1248-1262.

11. Böhmler, M.; Wang, Z.; Myaltsin, A.; Mews, A.; Hartschuh, A. *Angew. Chem. (Int. Ed.)* **2011**, *50,* 11536-11538.

12. Stöckle, R. M.; Suh, Y. S.; Deckert, V.; Zenobi, R. *Chem. Phys. Lett.* **2000**, *318,* 131-136.

13. Rauhut, N.; Engel, M.; Steiner, M.; Krupke, R.; Avouris, Ph.; Hartschuh, A. *ACS Nano* **2012**, *6,* 6416–6421.

14. Misewich, J. A.; Martel, R.; Avouris, Ph.; Tsang, J. C.; Heinze, S.; Tersoff, J. *Science* **2003**, *300,* 783-786.

15. Freitag, M.; Chen, J.; Tersoff, J.; Tsang, J. C.; Fu, Q.; Liu, J.; Avouris, Ph. *Phys. Rev. Lett.* **2004**, *93,* 076803.

16. Freitag, M.; Perebeinos, V.; Chen, J.; Stein, A.; Tsang, J. C.; Misewich, J. A.; Martel, R.; Avouris, Ph. *Nano Lett.* **2004**, *4,* 1063-1066.

17. Chen, J.; Perebeinos, V.; Freitag, M.; Tsang, J.; Fu, Q.; Liu, J.; Avouris, Ph. *Science* **2005**, *310,* 1171.

18. Marty, L.; Adam, E.; Albert, L.; Doyon, R.; Ménard, D.; Martel, R. *Phys. Rev. Lett.* **2006**, *96,* 136803.

19. Mann, D.; Kato, Y. K.; Kinkhabwala, A.; Pop, E.; Cao, J.; Wang, X.; Zhang, L.; Wang, Q.; Guo, J.; Dai, H. *Nature Nanotech.* **2007**, *2,* 33.

20. Xie, L; Farhat, H.; Son, H.; Zhang, J.; Dresselhaus, M. D.; Kong, J.; Liu, Z. *Nano Lett.* **2009**, *9*, 1747-1751.

21. Essig, S.; Marquardt, C. W.; Vijayaraghavan, A.; Ganzhorn, M.; Dehm, S.; Hennrich, F.; Ou, F.; Green, A. A.; Sciascia, C.; Bonaccorso, F.; Bohnen, K.-P.; Löhneysen, v. H.; Kappes, M. M.; Ajayan, P. M. ; Hersam, M. C. ; Ferrari, A. C. ; Krupke, R. *Nano Lett.* **2010**, *10*, 1589-1594.

22 Freitag, M.; Tsang, J. C.; Bol, A.; Yuan, D.; Liu, J.; Avouris, Ph. *Nano Lett.* **2007**, *7*, 2037-2042.

23. Lee, E. J. H.; Balasubramanian, K.; Dorfmüller, J.; Vogelgesang, R.; Fu, N.; Mews, A.; Burghard, M.; Kern. K. *Small* **2007**, *3*, 2038-2042.

24. Freitag, M.; Tsang, J. C.; Bol, A.; Yuan, D.; Liu, J.; Avouris, Ph. *Appl. Phys. Lett.* **2007**, *91,* 031101.

25. Ahn, Y. H.; Tsen, A. W.; Kim. B.; Park Y. W.; Park, J. *Nano Lett.* **2007**, *7*, 3320-3323.

26. Hofmann, M. S.; Glückert, J. T.; Noé, J.; Bourjau, C.; Dehmel, R.; Högele, A. *Nature Nanotech.* **2013**, *8,* 502-505.

27. Fuhrer, M. S.; Nygard, J.; Shih, L.; Forero, M.; Yoon, Y.-G.; Mazzoni, M. S. C.; Choi, H. J. Ihm, J.; Louie, S. G.; Zettl, A.; McEuen, P. L. *Science* **2000**, *288,* 494-497.

28. Lee, E. J. H.; Balasubramanian, K.; Burghard, M.; Kern, K. *Adv. Materials* **2009**, *21,* 2720-2724.

29. The detected EL image is a convolution of the actual EL source and the point spread function (PSF) of the near-field imaging system. The latter can be approximated by the cross section recorded perpendicular to the CNT due to the small CNT diameter. Assuming a Gaussian PSF and Gaussian shaped EL intensity distribution of the source, the measured signal width will be $w = (w^2(PSF) + w^2(EL\ source))^{1/2}$ where w(PSF) and w(EL source) are the width of the PSF and of the source, respectively. Since the measured width w is approximately equal to the measured w(PSF) with an absolute error of 20 nm, we conclude that the size of the EL source must be smaller than 20 nm.

30. Böhmler, M.; Hartmann, N.; Georgi, C.; Hennrich, F.; Green, A. A.; Hersam, M. C.; Hartschuh, A. *Opt. Express* **2010**, *18,* 16443.

31. Derycke, V.; Martel, R.; Appenzeller, J.; Avouris, Ph. *Appl. Phys. Lett.* **2002**, *80*, 2773.

32. Mueller, T.; Kinoshita, M.; Steiner, M.; Perebeinos, V.; Bol, A. A.; Farmer, D. B.; Avouris, Ph. *Nature Nanotech.* **2010**, *5*, 27.

33. Wang, S.; Zeng, Q.; Yang, L.; Zhang, Z.; Wang, Z.; Pei, T.; Ding, L.; Liang, X.; Gao, M.; Li, D.; Peng, L.-M. *Nano Lett.* **2011**, *11*, 23–29.

# Antenna-enhanced Optoelectronic Probing of Carbon Nanotubes

*Nina Mauser (1), Nicolai Hartmann (1), Matthias S. Hofmann (2), Julia Janik (1), Alexander Högele (2), Achim Hartschuh (1)\**

(1) Department Chemie und Center for NanoScience (CeNS), Ludwig-Maximilians-Universität, 81377 München, Germany

(2) Fakultät für Physik und Center for NanoScience (CeNS), Ludwig-Maximilians-Universität, 80539 München, Germany

\*E-mail: achim.hartschuh@cup.uni-muenchen.de

## Supporting Information

Supporting material is provided on the CNT network and its characterization by Raman spectroscopy in Figures S1 and S2 and Note 1.

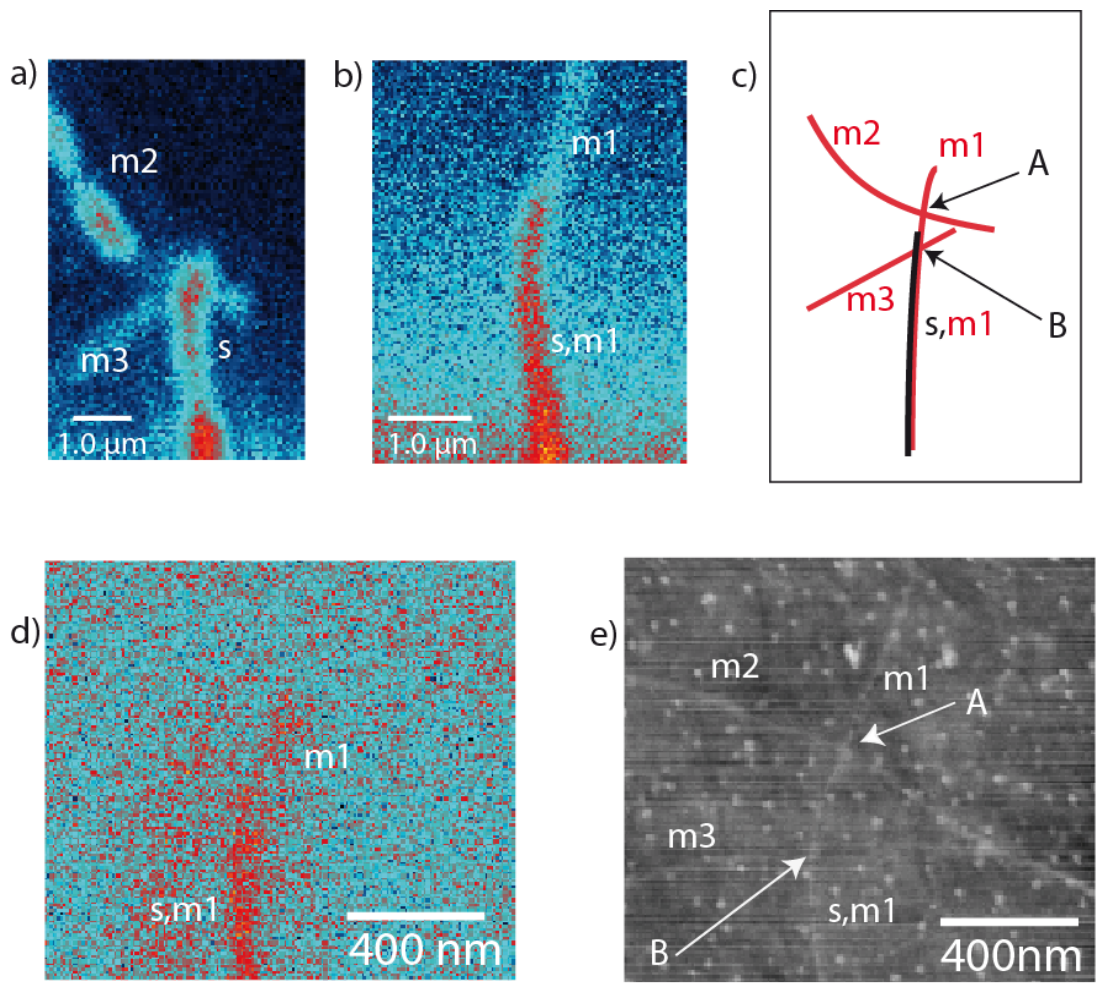

**Supporting Figure S1.** Overview of the CNT network a) Raman G band image with excitation wavelength 834 nm. b) Raman G band image with excitation wavelength 633 nm. c) Schematic drawing of the CNT network showing the involved CNTs. *m* stands for a metallic CNT, *s* for a semiconducting CNT as determined from the respective confocal Raman spectra in Figure S2 discussed in Note 1. *A* and *B* denote the two nanotube crossings referred to in the main text. d) Antenna-enhanced Raman G band image. e) Simultaneously taken topography image showing the two crossings *A* and *B*.

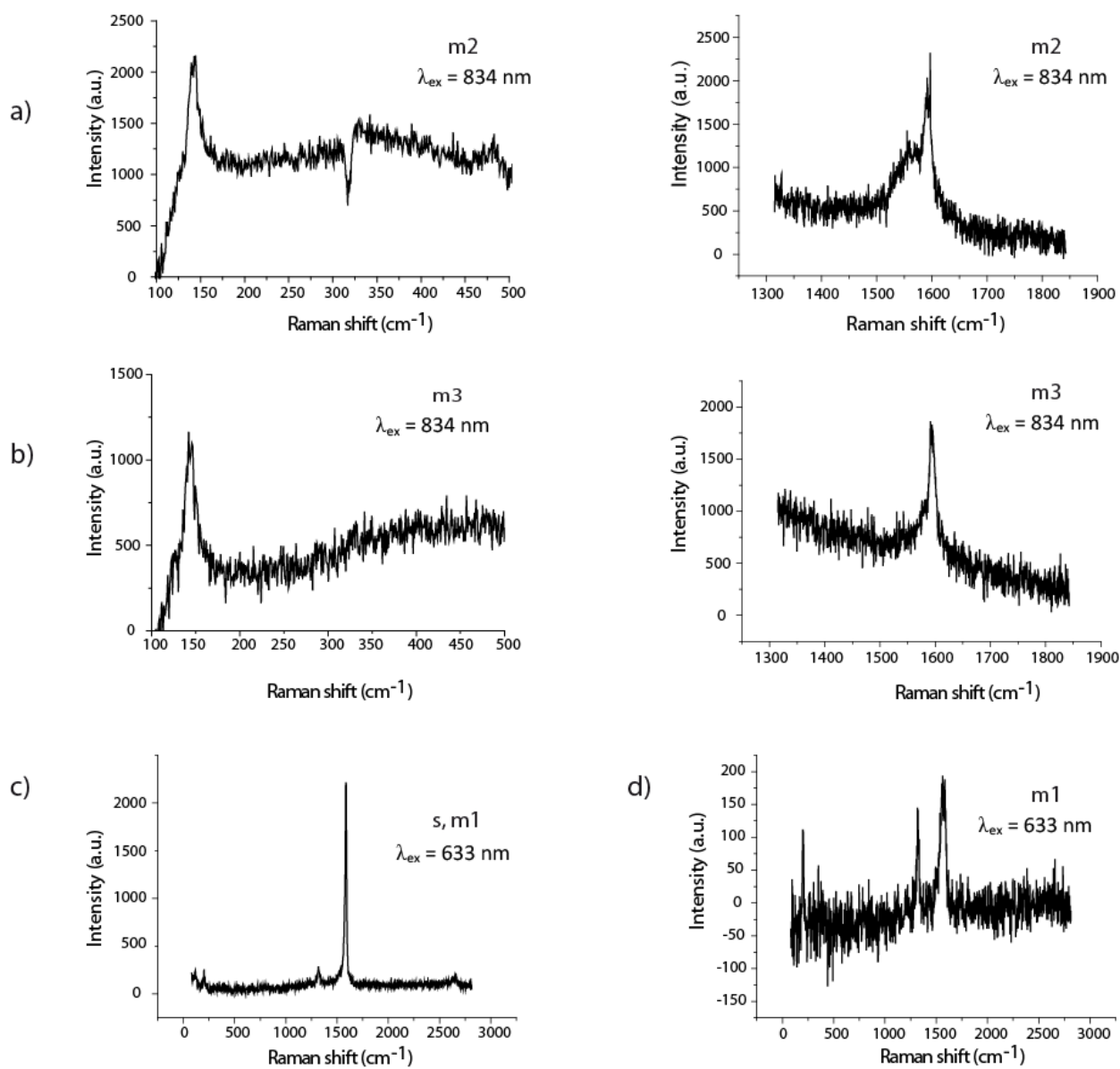

**Supporting Figure S2.** Raman spectra of the CNTs labeled in Figure S1 by *m1, m2, m3*. The shape of the G band at about 1600 cm$^{-1}$ together with the position of the radial breathing mode (RBM) peaks at low energies indicate the metallic character of these CNTs.

## Supporting Note 1

**Characterization of the CNT network via Raman microscopy**

With the help of supporting Figure S1 and S2 we can characterize the network according to the schematic drawing shown in Figure S1c:

*m1*: *m1* is a metallic CNT (FigureS2d). The RBM at 200 cm$^{-1}$ together with the excitation wavelength of 633 nm indicate that the CNT is metallic using the Kataura plot for CVD grown CNTs on quartz[1].

*m2* and *m3* are metallic CNTs. The RBM at 142 cm$^{-1}$ together with the excitation wavelength of 834 nm and the shape of the G band (FigureS2a and S2b) indicate that the CNTs are metallic[1].

*s*: The fact that this CNT shows a strong G band signal for both the excitation wavelengths 834 nm and 633 nm (FigureS1a and S1b) indicates that this CNT is semiconducting[1].

Crossing *A* is a crossing between two metallic CNTs as can be seen from the antenna-enhanced Raman image and the topography shown in Figure S1d and S1e. The Raman intensity increases only below crossing *A*, indicating that the semiconducting CNT s starts behind the crossing.

Crossing *B* involves at least the semiconducting CNT *s* and one metallic CNT (*m3*, eventually also *m1*).

REFERENCES:

1. Soares, J. S., Cancado L. G., Barros E. B. & Jorio, A. The Kataura plot for single wall carbon nanotubes on top of crystalline quartz. *Phys. Status Solidi B* 247, 2835–2837 (2010).